\documentclass{PoS}

\newcommand{\bea}{\begin{eqnarray}}
\newcommand{\eea}{\end{eqnarray}}

\def\spa#1.#2{\langle#1\,#2\rangle}
\def\spb#1.#2{[#1\,#2]}
\def\spab#1.#2.#3{\langle\mskip-1mu{#1} 
                  | #2 | {#3}]}

\def\spba#1.#2.#3{[\mskip-1mu{#1} 
                  | #2 | {#3}\rangle}

\def\spbb#1.#2.#3.#4{[\mskip-1mu{#1} 
                     | {#2} \ {#3} | {#4}]}

\def\spaa#1.#2.#3.#4{\langle\mskip-1mu{#1} 
                     | {#2} \ {#3} | {#4}\rangle}

\newbox\SlashedBox
\def\slashed#1{\setbox\SlashedBox=\hbox{#1}
\hbox to 0pt{\hbox to 1\wd\SlashedBox{\hfil/\hfil}\hss}#1}
\def\hboxtosizeof#1#2{\setbox\SlashedBox=\hbox{#1}
\hbox to 1\wd\SlashedBox{#2}}

\newbox\charbox
\newbox\slabox
\def\s#1{{      
        \setbox\charbox=\hbox{$#1$}
        \setbox\slabox=\hbox{$/$}
        \dimen\charbox=\ht\slabox
        \advance\dimen\charbox by -\dp\slabox
        \advance\dimen\charbox by -\ht\charbox
        \advance\dimen\charbox by \dp\charbox
        \divide\dimen\charbox by 2
        \raise-\dimen\charbox\hbox to \wd\charbox{\hss/\hss}
        \llap{$#1$}
}}

\title{Unitarity-Cuts, Stokes' Theorem and Berry's Phase}

\ShortTitle{Unitarity-Cuts, Stokes' Theorem and Berry's Phase}

\author{\speaker{Pierpaolo Mastrolia}\thanks{
In memory of my friend Thomas (Binoth), who reacted with  
his characteristic enthusiasm when I first shew him the results 
now collected in this manuscripts. 
I acknowledge Tania Robens for clarifying discussions.
My participation to the
conference was supported by HepTools.}\\
        Centro Studi e Ricerche ``E. Fermi'', 
        Piazza del Viminale 1, I-00184, Roma, Italy \\
        Dipartimento di Fisica, Universit\`a di Salerno, 
        Via Ponte don Melillo, I-84084 Fisciano, Italy \\
        Theory Department, CERN, CH-1211 Geneva 23, Switzerland \\
        E-mail: \email{pierpaolo.mastrolia@cern.ch}}

\abstract{
Two-particle unitarity-cuts of scattering amplitudes can be 
efficiently computed by applying Stokes' Theorem, in the fashion 
of the Generalised Cauchy Theorem.
Consequently, the Optical Theorem can be related to the Berry Phase,
showing how the imaginary part of arbitrary one-loop Feynman amplitudes 
can be interpreted as the flux of a complex 2-form.
}

\FullConference{RADCOR 2009 - 9th International Symposium on Radiative Corrections (Applications of Quantum Field Theory to Phenomenology) ,\\
		 October 25 - 30 2009\\
		 Ascona, Switzerland}

\begin{document}

\section{Introduction}

Unitarity and geometric phases are two ubiquitous properties
of physical systems.
The Berry phase is the phase acquired by a system when it is subjected 
to a cyclic evolution, resulting only from the geometrical properties 
of the path traversed in the parameter space because of anholonomy
\cite{Berry,Shapere:1989kp}.
Unitarity represents the probability conservation in particle 
scattering processes described by the unitary {\it scattering
operator}, $S$. 
The relation, $S = 1 + i \ T$,
between the $S$-operator and the {\it transition operator}, $T$,
leads to the Optical Theorem,
$-i (T - T^{\dagger}) =  T^{\dagger} T \ .$
The matrix elements of this equation between initial and final
states are expressed, in perturbation theory, 
in terms of Feynman diagrams. The evaluation of the right hand side requires
the insertion of a complete set of intermediate states.
Therefore, since $-i (T - T^{\dagger}) = 2 \ {\rm Im} T$,
the Optical Theorem yields the computation of the imaginary part of 
Feynman integrals from a sum of contributions from all possible
intermediate states. 

The Cutkosky-Veltman rules, implementing
the unitarity conditions, allow the calculation
of the discontinuity across a branch cut of an
arbitrary Feynman amplitude,
which corresponds to its imaginary part \cite{OldUnitarity}.
Accordingly, the imaginary
part of a given Feynman integral can be computed 
by evaluating the phase-space integral obtained 
by cutting two internal particles, which amounts 
to applying the on-shell conditions
and replacing their propagators by the corresponding 
$\delta$-function,
$(p^2 - m^2 + i0)^{-1} \to (2 \pi i) \ \delta^{(+)}(p^2 - m^2) .$
In later studies the problem of finding the discontinuity 
of a Feynman integral associated to a singularity was addressed in the language
of homology theory and differential forms \cite{homology}.

More recently multi-particle cuts have been combined 
with the use of complex momenta for on-shell 
internal particles into very efficient techniques, by-now known as
{\it unitarity-based methods}, to compute scattering amplitudes for arbitrary 
processes. 
These methods exploit two general properties of scattering amplitudes like
analyticity, granting that amplitudes are determined by 
their own singularity-structure, 
and unitarity, granting that the residues at the singular
points factorize into products of simpler amplitudes.
Unitarity and analyticity become tools for the quantitative 
determinaton of one-loop amplitudes \cite{Bern:1994zx}
when merged with the existence of an underlying representation of amplitudes
as a combination of basic scalar one-loop functions \cite{Passarino:1978jh}.
These functions, known as Master Integrals (MI's),
are $n$-point one-loop integrals, $I_n$ ($1\le n \le 4$),
with trivial numerator, equal to 1, characterised 
by external momenta and internal masses present in the denominator.  
In general, the fulfillment of multiple-cut conditions requires
loop momenta with complex components.
Since the loop momentum has four components, 
the effect of the cut-conditions is to
fix some of them according to the
number of the cuts. Any {\it quadruple}-cut \cite{Britto:2004nc}
fixes the loop-momentum completly, yielding the
algebraic determination of the coefficients 
of $I_n, (n\ge4)$;
the coefficient of 3-point functions, $I_3$,
are extracted from {\it triple}-cut 
\cite{MastroliaTriple,Binoth:2007ca,FordeTriBub,BjerrumBohr:2007vu,Kilgore:2007qr,Badger:2008cm};
the evaluation of {\it double}-cut 
\cite{Britto:2005ha,Britto:2006sj,ABFKM,FordeTriBub,Britto:2007tt,Kilgore:2007qr,Britto:2008vq,Britto:2008sw,Badger:2008cm}
is necessary for extracting the coefficient of 2-point
functions, $I_2$; and 
finally, in processes involving massive particles,
the coefficients of 1-point functions, $I_1$, are detected by 
{\it single}-cut \cite{Kilgore:2007qr,NigelGlover:2008ur,Britto:2009wz}.
In cases where fewer than four denominators are cut, the loop momentum
is not frozen: the free-components are left over as phase-space 
integration variables.

In \cite{Mastrolia:2009dr,Mastrolia:2009rk},
I introduced a novel efficient method for the 
analytic evaluation of the coefficients of one-loop 2-point 
functions {\it via} double-cuts.
Spun-off from the spinor-integration technique 
\cite{Britto:2005ha,Britto:2006sj,ABFKM},
that method is an 
application of Stokes' Theorem.
Due to a special decomposition of the loop-momentum,
the double-cut phase-space integral is written
as parametric integration of rational function 
in two complex-conjugated variables.
By applying Stokes' Theorem, the integration is carried on in 
two simple steps:
an indefinite integration in one variable, followed by Cauchy's
Residue Theorem in the conjugated one. 

The coefficients of the 2-point scalar functions, being proportional 
to the rational term of the double-cut, can be directly 
extracted from the indefinite integration by Hermite Polynomial Reduction.

\section{Double-Cut}
\label{sec:integration}

The two-particle Lorentz invariant phase-space (LIPS) in the $K^2$-channel
is defined as,
\bea
\int d^4\Phi = 
\int d^4 \ell_1 
\ \delta^{(\!+\!)}(\ell_1^2 \!-\! m_1^2) 
\ \delta^{(\!+\!)}((\ell_1-K)^2 \!-\! m_2^2) \ , 
\label{def:phi4}
\eea
where $K^\mu$ is the total momentum across the cut.
We introduce a suitable parametrization
for $\ell_1^\mu$ \cite{Mastrolia:2009dr,ABFKM}, 
in terms of four massless momenta,
which is a solution of the two on-shell conditions,
$\ell_1^2 = m_1^2$ and $ (\ell_1-K)^2 = m_2^2$,
\bea
\ell_1^\mu = {1 - 2 \rho \over 1 + z \bar{z}} \Big(
p^\mu + 
z \bar{z} \ q^\mu
+  z \ \epsilon_{+}^\mu   
+  \bar{z} \ \epsilon_{-}^\mu \Big) + \rho K^\mu \ ,
\label{def:loopdeco}
\eea
where 
$p_\mu$ and $q_\mu$ 
are two massless momenta with the requirements,
\bea
p_\mu + q_\mu = K_\mu \ , \qquad 
p^2=q^2=0 \ ,  \qquad
2 \ p\cdot q = 2 \ p\cdot K = 2 \ q\cdot K \equiv K^2 \ ;
\label{def:specialpq}
\eea
the vectors $\epsilon_{+}^\mu$ and 
$\epsilon_{-}^\mu$ 
are orthogonal to both
$p^\mu$ and $q^\mu$, with the following properties \footnote{
In terms of spinor variables 
that are associated to massless momenta, 
we can define
$p^\mu = (1/2){\spab p.\gamma^\mu.p }$ and 
$q^\mu = (1/2){\spab q.\gamma^\mu.q }$,
hence $\epsilon_{+}^\mu = (1/2){\spab q.\gamma^\mu.p }$
and $\epsilon_{-}^\mu = (1/2){\spab p.\gamma^\mu.q }$.},
\bea
\epsilon_{+}^2 = \epsilon_{-}^2 = 0 = 
\epsilon_{\pm}\cdot p = \epsilon_{\pm}\cdot q \ , \qquad
2 \ \epsilon_{+} \cdot \epsilon_{-} = - K^2 \ . 
\eea
The pseudo-threshold
$
\rho = (K^2 + m_1^2 - m_2^2 - \sqrt{\lambda})/ (2 K^2) \ ,
$
with 
$
\lambda =
(K^2)^2 + (m_1^2)^2 + (m_2^2)^2 
- 2 K^2 m_1^2
- 2 K^2 m_2^2
- 2 m_1^2 m_2^2 \ , $
depends only on the kinematics. 
The complex conjugated variables $z$ and $\bar{z}$ parametrize
the degrees of freedom left over by the cut-conditions.
Because of (\ref{def:loopdeco}),
the LIPS in (\ref{def:phi4}) reduces to the remarkable expression,
\bea
\int d^4 \Phi = (1-2\rho)
\int \!\!\!\! \int
{dz \wedge d\bar{z} \over (1 + z \bar{z})^2} \ .
\label{eq:novelphi4}
\eea

\begin{figure}[t]
\begin{center}
\vspace*{-1cm}
\includegraphics[scale=0.15]{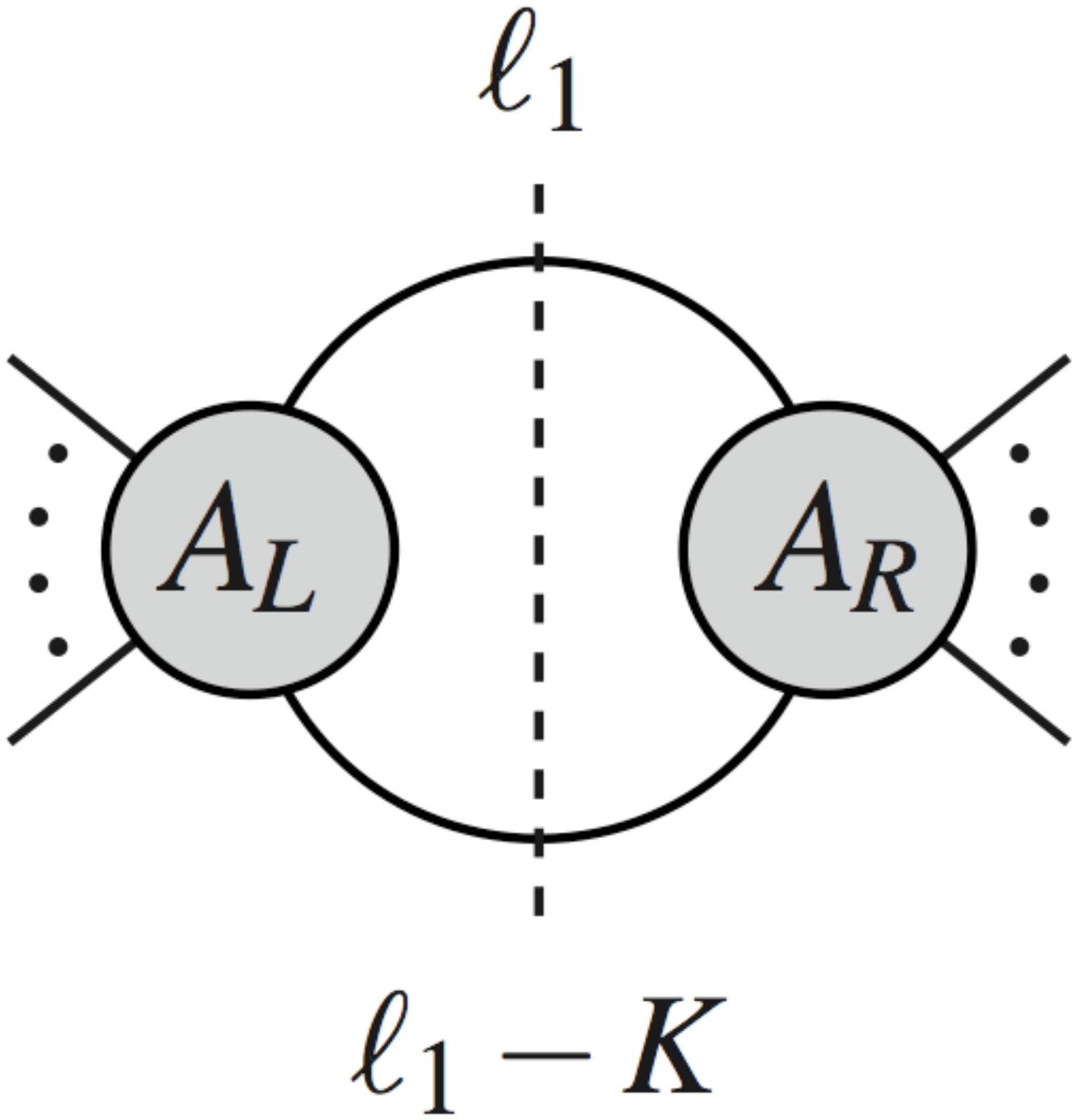}
\vspace*{-1.0cm}
\caption{Double-cut of one-loop amplitude in the $K^2$-channel.}
\label{fig:doublecut}
\end{center}
\end{figure}

The double-cut of a generic $n$-point amplitude
in the $K^2$-channel is defined as
\bea
\Delta \equiv 
\int d^4\Phi \  
A^{\rm tree}_L(\ell_1) \ 
A^{\rm tree}_R(\ell_1) \ ,
\eea
where $A^{\rm tree}_{L,R}$ are the tree-level amplitudes
sitting at the two sides of the cut, see Fig.\ref{fig:doublecut}.
By using (\ref{eq:novelphi4}) for the LIPS, 
and (\ref{def:loopdeco}) for the loop-momentum $\ell_1^\mu$,
one has,
\bea
\Delta =
(1-2\rho) \!\!
\int \!\!\!\! \int
dz \wedge d\bar{z} \ 
{ A^{\rm tree}_L( z, \bar{z}) \ 
A^{\rm tree}_R( z, \bar{z}) \over (1 + z \bar{z})^2} \ , \quad 
\label{def:zbarzInt}
\eea
where the tree-amplitudes $A^{\rm tree}_L$ and $A^{\rm tree}_R$ are rational
in $z$ and $\bar{z}$. Since $\rho$ is independent of $z$ and $\bar{z}$,
its presence in the integrand is understood.
By applying a special version of the
so called {\it Generalised Cauchy Formula} 
also known as the {\it Cauchy-Pompeiu Formula} \cite{CauchyPompeiu}, 
one can write the two-fold integration 
in $z$- and $\bar{z}$-variables
appearing in Eq.(\ref{def:zbarzInt}) simply as 
a convolution of an unbounded $\bar{z}$-integral 
and a contour $z$-integral\footnote{
The roles of $z$ and $\bar{z}$ can be equivalently exchanged.
} \cite{Mastrolia:2009dr},
\bea
\Delta =
(1-2\rho) 
 \oint d z \!\! \int \!\! d\bar{z} \ 
{ A^{\rm tree}_L(z, \bar{z}) \ 
A^{\rm tree}_R(z, \bar{z}) \over (1 + z \bar{z})^2} \ 
 \ ,
\label{def:practicalzbarzInt}
\eea 
where 
the product $A^{\rm tree}_L A^{\rm tree}_R $ is a rational function
of $z$ and $\bar{z}$, and the integration contour has to be chosen 
as enclosing all the complex $z$-poles.
The equivalence of Eq.(\ref{def:zbarzInt}) and Eq.(\ref{def:practicalzbarzInt})
is due to Stokes' Theorem \cite{Mastrolia:2009dr}. Accordingly, 
the double-cut $\Delta$ in (\ref{def:zbarzInt}) is the flux of a 2-form,
corresponding to an integral over the 
complex tangent bundle of the Riemann sphere\footnote{
In \cite{Mastrolia:2009dr} it has been shown that the double-cut of the scalar
2-point function, $\Delta I_2$ = $\int d^4\Phi$ amounts to the integral 
$\int \!\!\! \int \Omega = - 2 \pi i$. This result corresponds
to the integration of the first Chern class, $(i/\pi) \int \!\!\! \int \Omega = 2$.
},
where the curvature 2-form is defined as 
$\Omega = 1/(1 + |z|^2)^2 \ dz \wedge d\bar{z} \ .$

\subsection{Coefficient of the 2-point function}

The formula in Eq.(\ref{def:practicalzbarzInt}) can be intrgrated 
straightforwardly in two steps.
To begin with the integration, 
we find a primitive with respect
to $\bar{z}$, say $F$, by keeping $z$ as independent variable,
\bea
F(z,\bar{z}) =
\int d\bar{z} \ { A^{\rm tree}_L(z, \bar{z}) \ 
A^{\rm tree}_R(z, \bar{z}) \over (1 + z \bar{z})^2}  \ , 
\label{eq:primitive}
\eea
so that $\Delta$ becomes,
\bea
\Delta = 
(1-2\rho)
\oint dz \ F(z, \bar{z}) \ .
\label{eq:dummyDelta}
\eea
Since $F$ is the primitive of a rational function, its general
form can only contain two types of terms: 
a rational term and a logarithimc one,
\bea
F(z,\bar{z}) = F^{\rm rat}(z,\bar{z}) + F^{\rm log}(z,\bar{z}) \ .
\label{eq:EFFdef}
\eea
%
The coefficient of a 2-point function in the $K^2$-channel
will appear in $\Delta^{\rm rat}$, namely the result of the Residue Theorem 
in $z$ applied only to $F^{\rm rat}$.
The choice of $p$ and $q$ specified in Eqs.(\ref{def:specialpq})
grants that there exists a pole at $z=0$ associated to the 
2-point function in the $K^2$-channel, $I_2(K^2)$;
while the reduction of higher-point functions that have $I_2(K^2)$ 
as subdiagram can generate poles at finite $z$-values.  
The Residue Theorem has to be applied by reading all the residues
in $z$, and substituting the corresponding complex-conjugate values
where $\bar{z}$ appears.
Since it can be shown \cite{Mastrolia:2009dr} that the double-cut 
of the 2-point scalar function in the $K^2$-channel amounts to 
$
\Delta I_2 = - 2 \pi i \  (1-2\rho) ,
$
the coefficient of the 2-point function can be finally extracted from 
the ratio,
\bea
c_2 = {\Delta^{\rm rat} \over \Delta I_2} = - \Big(
{\rm Res}_{z = 0} \ F^{\rm rat}(z, \bar{z}) + 
{\rm Res}_{z \ne 0} \ F^{\rm rat}(z, \bar{z}) \Big)\ ,
\eea
which will depend on $\rho$.

Recently, the method just described has been succesfully applied to 
the completion of the analytic calculation of the one-loop virtual 
corrections to $H+2$jets {\it via} gluon fusion \cite{Badger:2009hw,Badger:2009vh}.\footnote{See also S. Badger, and C. Williams in these proceedings.}

\section{Optical Theorem and Berry's Phase}

In \cite{Mastrolia:2009rk} the following observation was made.
In the double-cut integral (\ref{def:zbarzInt}),
we did not make any assumptions on the tree-level amplitudes sewn along 
the cut, thus providing a general framework to the integration method developed in \cite{Mastrolia:2009dr}.
If we now choose $A_L^{\rm tree} = A_{m \to 2}^{*, \rm tree}$,
that is the conjugate scattering amplitude of a process $m \to 2$,
and $A_R^{\rm tree} = A_{n \to 2}^{\rm tree}$, that is the amplitude
of a process $n \to 2$, 
then $\Delta$ reads,
\bea
\Delta 
=
\int d^4\Phi \ 
 A^{*, \rm tree}_{m \to 2} \ A^{\rm tree}_{n \to 2}  
= 
- i \Big[ A^{\rm one-loop}_{n \to m} 
         - A^{*, \rm one-loop }_{m \to n}
    \Big] = 
2 \ {\rm Im}\Big\{ A^{\rm one-loop}_{n \to m} \Big\} \ ,
\label{eq:OpticalTheorem}
\eea
which is the definition of the 
two-particle discontinuity of the one-loop amplitude 
$A^{\rm one-loop}_{n \to m}$ across the branch cut in the  $K^2$-channel, 
corresponding to the field-theoretic version
of the Optical Theorem for one-loop Feynman amplitudes.
On the other side, because of Stokes' Theorem in (\ref{def:zbarzInt}, \ref{def:practicalzbarzInt}), one has,
\bea
\Delta =
(1-2\rho) 
\int \!\!\!\! \int \!\!
dz \wedge d\bar{z} \ 
{   A^{*, \rm tree}_{m \to 2} \ A^{\rm tree}_{n \to 2} 
\over (1 + z \bar{z})^2} \ 
= 
(1-2\rho) 
\oint dz \!\!
\int \!\! d\bar{z} \ 
{ A^{*, \rm tree}_{m \to 2} \ A^{\rm tree}_{n \to 2}
 \over (1 + z \bar{z})^2} \ ,
\label{eq:OpticalTheoremGeometric}
\eea
which provides a geometrical interpretation
of the imaginary part of one-loop scattering amplitudes,
as a flux of a complex 2-form through a surface bounded by the contour
of the $z$-integral (the contour should enclose all the poles in $z$ exposed
in the integrand after the integration in $\bar{z}$ \cite{Mastrolia:2009dr}).

Given the equivalence of (\ref{eq:OpticalTheorem}) and 
(\ref{eq:OpticalTheoremGeometric}), 
a correspondence between
the imaginary part of scattering amplitudes and the anholonomy of 
Berry's phase does emerge,
since the latter is indeed defined
as the flux of a 2-form in presence of curved 
space \cite{Berry,Shapere:1989kp}. 
In this context, one could establish a parallel description
between the Aharonov-B\"ohm (AB) effect \cite{Aharonov:1959fk} 
and the double-cut of one-loop Feynman integrals. 
Accordingly, let us follow the evolution in Fig.\ref{fig:doublecut} from the 
left to the right. The two particles produced in the $A_L$-scattering, going
around the loop and initiating the $A_R$-process, 
at the  $A_R$-interaction point would experience 
a phase-shift due to the non-trivial geometry in effective 
momentum space induced by the on-shell conditions.
As in the AB-effect, the anholonomy phase-shift is a consequence 
of Stokes' Theorem, and here it corresponds to the imaginary part of
the one-loop Feynman amplitude.


\begin{thebibliography}{99}
\bibitem{Berry}
  M.~V.~Berry,
  Proc.\ Roy.\ Soc.\ Lond.\  A {\bf 392} (1984) 45.

\bibitem{Shapere:1989kp}
  A.~D.~Shapere and F.~Wilczek (ed.),
  Adv.\ Ser.\ Math.\ Phys.\  {\bf 5} (1989) 1.

\bibitem{OldUnitarity}{
L.~D.~Landau,
Nucl.\ Phys.\  {\bf 13}, 181 (1959). 
%
R.~E.~Cutkosky,
J.\ Math.\ Phys.\  {\bf 1}, 429 (1960). 
%
R.~J.~Eden,  P.~V.~Landshoff, D.~I.~Olive, J.~C.~Polkinghorne,
{\it The Analytic S Matrix}, Cambridge University Press, 1966. 
%
  M.~J.~G.~Veltman,
  Physica {\bf 29} (1963) 186. 
%
E. Remiddi,
 Helv. Phys. Acta {\bf 54} (1982) 364.
}

\bibitem{homology}
 R. C. Hwa, and V. L. Teplitz,
 Mathematical Physics Monographs,
 W. A. Benjamin Inc., 1966.

\bibitem{Bern:1994zx}
  Z.~Bern, L.~J.~Dixon, D.~C.~Dunbar and D.~A.~Kosower,
  Nucl.\ Phys.\  B {\bf 425} (1994) 217;
  Nucl.\ Phys.\ B {\bf 435} (1995) 59.


\bibitem{Passarino:1978jh}
  G.~Passarino and M.~J.~G.~Veltman,
  Nucl.\ Phys.\  B {\bf 160} (1979) 151.




\bibitem{Ossola:2006us}
  G.~Ossola, C.~G.~Papadopoulos and R.~Pittau,
  Nucl.\ Phys.\  B {\bf 763}, 147 (2007);
%
  JHEP {\bf 0707}, 085 (2007).


\bibitem{Ellis:2007br}
  R.~K.~Ellis, W.~T.~Giele and Z.~Kunszt,
  JHEP {\bf 0803} (2008) 003.

\bibitem{Giele:2008ve}
  W.~T.~Giele, Z.~Kunszt and K.~Melnikov,
  JHEP {\bf 0804} (2008) 049.


\bibitem{Berger:2008sj}
  C.~F.~Berger {\it et al.},
  Phys.\ Rev.\  D {\bf 78} (2008) 036003.






\bibitem{Britto:2004nc}
  R.~Britto, F.~Cachazo and B.~Feng,
  Nucl.\ Phys.\ B {\bf 725}, 275 (2005).

\bibitem{MastroliaTriple}
P.~Mastrolia,
Phys.\ Lett.\  B {\bf 644}, 272 (2007).

\bibitem{Binoth:2007ca}
  T.~Binoth, G.~Heinrich, T.~Gehrmann and P.~Mastrolia,
  Phys.\ Lett.\  B {\bf 649}, 422 (2007).



\bibitem{FordeTriBub}
  D.~Forde,
  Phys.\ Rev.\  D {\bf 75}, 125019 (2007).


\bibitem{BjerrumBohr:2007vu}
  N.~E.~J.~Bjerrum-Bohr, D.~C.~Dunbar and W.~B.~Perkins,
  JHEP {\bf 0804}, 038 (2008).

\bibitem{Kilgore:2007qr}
  W.~B.~Kilgore,
  arXiv:0711.5015 [hep-ph].

\bibitem{Badger:2008cm}
  S.~D.~Badger,
  JHEP {\bf 0901} (2009) 049.


\bibitem{Britto:2005ha}
  R.~Britto, E.~Buchbinder, F.~Cachazo and B.~Feng,
  Phys.\ Rev.\ D {\bf 72}, 065012 (2005).

\bibitem{Britto:2006sj}
  R.~Britto, B.~Feng and P.~Mastrolia,
  Phys.\ Rev.\ D {\bf 73}, 105004 (2006).

\bibitem{ABFKM}{
C.~Anastasiou, R.~Britto, B.~Feng, Z.~Kunszt and P.~Mastrolia,
Phys.\ Lett.\  B {\bf 645}, 213 (2007);
%
JHEP {\bf 0703}, 111 (2007).
}


\bibitem{Britto:2007tt}{
%
  R.~Britto and B.~Feng,
  Phys.\ Rev.\  D {\bf 75}, 105006 (2007);
%
  JHEP {\bf 0802}, 095 (2008).
}

\bibitem{Britto:2008vq}
  R.~Britto, B.~Feng and P.~Mastrolia,
  Phys.\ Rev.\  D {\bf 78} (2008) 025031.

\bibitem{Britto:2008sw}
  R.~Britto, B.~Feng and G.~Yang,
  JHEP {\bf 0809} (2008) 089.





\bibitem{NigelGlover:2008ur}
  E.W.N. Glover and C.~Williams,
  JHEP {\bf 0812} (2008) 067.

\bibitem{Britto:2009wz}
  R.~Britto and B.~Feng,
  Phys.\ Lett.\  B {\bf 681}, 376 (2009).


\bibitem{Mastrolia:2009dr}
  P.~Mastrolia,
  Phys.\ Lett.\  B {\bf 678} (2009) 246.

\bibitem{Mastrolia:2009rk}
  P.~Mastrolia,
  to appear in Lett. Math. Phys.,
  arXiv:0906.3789. 

\bibitem{CauchyPompeiu}
 M. J. Ablowitz, and A. S. Fokas,
 {\it Complex Variables},
 Cambridge Texts in Applied Mathematics, 2003, 2nd Edition.

\bibitem{Badger:2009hw}
  S.~Badger, E.~W.~Nigel Glover, P.~Mastrolia and C.~Williams,
  to appear in JHEP, arXiv:0909.4475. 

\bibitem{Badger:2009vh}
  S.~Badger, J.~M.~Campbell, R.~K.~Ellis and C.~Williams,
  JHEP {\bf 0912} (2009) 035.


\bibitem{Aharonov:1959fk}
  Y.~Aharonov and D.~Bohm,
  Phys.\ Rev.\  {\bf 115} (1959) 485.

\end{thebibliography}
\end{document}